\documentclass[prl,twocolumn,showkeys,floatfix]{revtex4}
\usepackage{graphicx,amssymb,color}

\def\be{\begin{eqnarray} &&}
\def\nonu{\nonumber \\ &&}

\def\ee{\end{eqnarray}}
\def\psla{\slash \! \! \!}
\newcommand{\eqref}[1]{(\ref{#1})}

\begin{document}
\title{Advances in solving the two-fermion homogeneous Bethe-Salpeter equation 
in Minkowski space
}
\author{W. de Paula$^a$, T.   Frederico$^a$, G. Salm\`e$^b$ and   M. Viviani$^c$ 
 }
\affiliation{$^a$ Dep. de F\'\i sica, Instituto Tecnol\'ogico da Aeron\'autica,
Centro T\'ecnico Aeroespacial, 12.228-900 S\~ao Jos\'e dos
Campos, S\~ao Paulo, Brazil\\
$^b$Istituto  Nazionale di Fisica Nucleare, Sezione di Roma, P.le A. Moro 2,
 I-00185 Rome, Italy \\ 
$^c$Istituto  Nazionale di Fisica Nucleare, Sezione di Pisa, 
Largo Pontecorvo 3, 56100, Pisa, Italy }
\date{Received: date }
\begin{abstract}
    Actual solutions of the Bethe-Salpeter
 equation  for a two-fermion  bound system  are  becoming available directly 
 in Minkowski space, by virtue of   a  novel technique,
 based on the so-called Nakanishi integral representation of the Bethe-Salpeter
  amplitude and  improved by expressing the relevant momenta through
  light-front components,  i.e. $k^\pm=k^0\pm k^3$.
 We solve a crucial problem that widens the applicability of the method to real situations 
 by providing an  analytically exact treatment of the singularities
plaguing   the two-fermion problem 
 in Minkowski space,  irrespective of the complexity of the irreducible Bethe-Salpeter kernel.  
  This  paves the way for feasible numerical
  investigations of relativistic 
    composite systems, with  any spin degrees of freedom.
  We present  a thorough comparison with existing numerical results, evaluated 
 in both Minkowski 
and Euclidean space, fully 
corroborating our analytical  treatment,   as well as fresh light-front 
amplitudes illustrating the potentiality
of non perturbative  calculations performed directly in Minkowski space.
\end{abstract}
\maketitle
 To solve the bound-state problem in relativistic  field theory, directly
  in Minkowski space,
 is still a challenge, and to cope with it by means of a viable tool  is of 
 wide interest in  many areas, from  condensed matter to nuclear and hadron
  physics, whenever dynamical observables, like momentum distributions, are
  needed. In view of this, integral equations represent a non perturbative
  framework to be explored.
 
More than half a century ago, in a seminal work \cite{BS51} Salpeter and Bethe  
 presented a dynamical equation for describing bound systems within the  
 relativistic  field theory.
In the subsequent years, there has been a large number of
applications of their integral equation, but mainly adopting Euclidean
variables or effective reduction to a 3D space.  More recently,   
a method based on  the so-called Nakanishi integral  representation 
(NIR) of the
Bethe-Salpeter amplitude (see, e.g.,  Ref. \cite{FSV1} 
 and references  therein), has
allowed to make substantial steps forward in obtaining accurate 
numerical solutions of the  actual Bethe-Salpeter equation (BSE). 
 With massive-boson exchanges, it has been investigated: 
(i) two-scalar   bound and 
zero-energy states
\cite{Kusaka,CK2006,FSV2,Tomio2016,FSV3}       
  as well as  two-fermion  ground states \cite{CK2010},  
   with a  ladder kernel,  governing, as well-known,  the tail
  of the momentum distributions; (ii)   
   a two-scalar system, with  a cross-ladder kernel
 \cite{CK2006b}.

In this Letter, we first  present   the formally exact integration of the 
singularities that prevent a straightforward
application of the NIR for solving the two-fermion ladder BSE in  Minkowski 
space,  as it was accomplished in the case  of  two-scalar
systems \cite{CK2006,FSV2,Tomio2016,FSV3}.  Then,  after exactly
transforming   BSE  in a coupled eigen-equation
system, we compare our
  eigenvalues  with both (i)
the ones  still obtained  in Minkowski space \cite{CK2010}, but introducing
 an auxiliary smoothing
function, and (ii)
 outcomes  in Euclidean space \cite{Dork}.  Our analysis, though  in ladder
  approximation,
 is fully able to address
  a relevant issue for hadron physics, i.e.  the tail of momentum distributions 
  of a fermionic system\cite{Ji2003}. For illustration,  the needed
  amplitudes are presented. Moreover, we
   establish a simple
 counting rule for the  singularities appearing when    
  constituents with higher spin are considered, irrespective of the
  kernel complexity.   Fortunately, our  numerical 
 procedure 
 allows us
 to face  with such generalizations. 

 The homogeneous BSE  
for a  two-fermion   system, as given in Ref. \cite{CK2010}, reads
\be
\Phi(k,p)=~S(p/2+k)~\int d^4k' ~F^2(k-k')i{\cal K}(k,k')
\nonu \times~~\Gamma_1~\Phi(k',p)~\bar\Gamma_2~
S(k-p/2)
\label{bse}\ee
where $\Phi(k,p)$ is the BS amplitude, $p/2\pm k$ the  four-momenta 
of the
off-mass-shell fermionic constituents, $p^2=M^2$  the   square mass of the 
system, and $$S(q)=i{\psla q  +m\over  q^2-m^2+i\epsilon}$$ the Dirac
 propagator. In Ref. \cite{CK2010}, $\Gamma_1=\Gamma_2$  was taken equal to
 $1,~\gamma_5$ and $\gamma^\mu$,
corresponding to 
 scalar, pseudoscalar and vector Dirac structure of the interaction vertexes 
 between
the constituents and the exchanged boson, while $
F(k-k')= {(\mu^2-\Lambda^2)/ [(k-k')^2-\Lambda^2 +i\epsilon]}
$ is a  vertex form factor.  The dimensionless 
coupling constant, $g$, and  the
momentum-dependent part of the  exchanged-boson propagator are contained
in $i{\cal K}$.  In ladder approximation,  we will consider:  (i) scalar and
pseudoscalar kernels ${\cal
K}=\pm~g^2/[(k-k')^2-\mu^2+i\epsilon]$ (plus for the first case
and  minus  for the second one); (ii)  a massless vector exchange, 
 i.e. ${\cal
K}^{\mu\nu}=~g^2~g^{\mu\nu}/[(k-k')^2+i\epsilon]$.
In Eq. \eqref{bse}, 
 $\bar \Gamma_2= C~\Gamma^T_2~C^{-1}$, where  $C$ is the charge conjugation and $T$
indicates the transpose.

As in \cite{CK2010}, our formal analysis  focuses on   two fermions 
  in a $J^\pi=0^+$ state.  In this case, the BS amplitude  is decomposed in four terms
\be
\Phi(k,p)= S_1 ~\phi_1(k,p)+S_2 ~\phi_2(k,p)+S_3 ~\phi_3(k,p)
\nonu +S_4 ~\phi_4(k,p)
\label{bsa}\ee
where $\phi_i$ are unknown scalar functions with well-defined symmetry
under the exchange $1 \to 2$, dictated by the symmetry of both $\Phi(k,p)$
and  the matrices $S_i$.   A suitable choice of them is the following~
\cite{CK2010}
\be
S_{1} = \gamma_5~~, \quad 
S_{2} = {\psla p\over M}  ~\gamma_5~~, \quad 
S_{3} = {k \cdot p \over M^3}  \psla p ~\gamma_5 - {1\over M} \psla k 
\gamma_5~~, \nonu
S_{4} = {i \over M^2} \sigma^{\mu\nu}  p_{\mu} k_{\nu} ~\gamma_5 ~~.
\label{S_structure}
\ee
{ where $S_i$ are orthogonal each other,} i.e.
$Tr\Bigl[S_i~S_j\Bigr]={\cal N}_i(k,p)~\delta_{ij}$, so that one can transform
Eq. \eqref{bse} for $J^\pi=0^+$ into a system of four coupled integral 
equations,  viz
 \be
\phi_i(k,p)= ig^2  \sum_{j}~\int {d^4k''
\over (2 \pi)^4}~{\phi_j(k'',p)\over (k-k'' )^2 -\mu^2+i\epsilon}\nonu \times~{ c_{ij}(k,k'',p)~F^2(k-k'')\over \Bigl[({p\over
2}+k)^2-m^2+i\epsilon\Bigr]~ \Bigl[({p\over
2}-k)^2-m^2+i\epsilon\Bigr]} 
\label{coupls1}\ee
with $i,j=1,~2,~3,~4$.  The  coefficients $c_{ij}(k,k'',p)=Tr\{S_i(k) 
(\psla p/2+\psla k+m) \Gamma_1 ~S_j(k'')\bar \Gamma_2(\psla p/2-\psla k-m)\}/{\cal N}_i(k,p)$
are explicitly given in Ref.~\cite{CK2010}  (a part a minor misprint, see
\cite{dFSV1} for details), for all the three couplings.
{ Notably,} the numerator of each $c_{ij}(k,k'',p)$ can
contain the third power of the four-momentum $k$, at the most.

In complete analogy with the two-scalar interacting system,
where only one amplitude is present 
\cite{Kusaka,CK2006,FSV2,Tomio2016,FSV3,CK2010,CK2006b}, one can introduce NIR  for each
amplitudes $\phi_i$, viz
\be
\phi_i(k,p)=\int_{-1}^1 dz'\int_0^\infty d\gamma' { g_i(\gamma',z';\kappa^2) \over 
\left[{k}^2+z' p\cdot k -\gamma'-\kappa^2+i\epsilon\right]^n}\nonu
\label{nakaw}\ee
where $n=3$ (see, e.g., the discussion  in Refs.
\cite{CK2006,FSV1}), $\kappa^2=m^2-M^2/4$ and $g_i(\gamma',z';\kappa^2)$ are unknown real functions, 
 called the Nakanishi weight
functions, to be numerically determined through the solutions of the
eigen-problem formally generated  after inserting  the above
  NIR in the BSE. The valuable second ingredient,
 that  greatly facilitates to get   numerical
solutions of the BSE, is represented by the use of light-front (LF)  components for the involved
momenta, i.e. 
$q^\pm=q^0\pm q^3$ and ${\bf
q}_\perp$. As it is well-known the LF variables allow  to simplify the analytic
integrations one meets, since one can  translate a double pole in
$k^0$ in two single poles in $k^-$ and $k^+$, obtaining great benefits in
the actual calculations 
(see, e.g., the discussions in \cite{FSV1,FSV2,Tomio2016,FSV3,dFSV1}).

In what follows,  we  fully exploit the advantages offered by the LF
 formalism, having the challenge
 to face with  singularities in $k^-$, 
called end-point
singularities. Within the LF quantization (see Ref. \cite{Yan}), they are 
related to the so-called instantaneous terms (in LF time) and  usually 
discussed in a perturbative regime, while,
 this
time, the framework is 
a non perturbative one. 

 Differently from  Ref. 
\cite{CK2010}, we  integrate both sides of Eq. \eqref{coupls1} on $k^-$, 
 getting (see Ref.
\cite{dFSV1} for details)
\be
\psi_{i}(\gamma,z)= g^2
\sum_{j}~\int_{-1}^1 dz' \int_0^\infty d\gamma' ~
g_j(\gamma',z';\kappa^2) \nonu \times ~
{\cal L}_{ij}(\gamma,z,\gamma',z';p) 
\label{coupls3}\ee
where { $\psi_{i}(\gamma,z)$ are the LF projection of the 
amplitudes $\phi_i$ and are given by }(see  Ref. \cite{FSV2})
\be
\psi_{i}(\gamma,z)=\int {dk^-\over 2\pi}~\phi_i(k,p)
=  -{i \over M} \nonu \times\int_0^\infty d\gamma' 
{ g_i(\gamma',z;\kappa^2) \over \left[\gamma + \gamma' + m^2 z^2 +
 (1-z^2)\kappa^2-i\epsilon\right]^2}
~,
\label{coupls4}\ee
with $z=-2
k^+/M$, $\gamma=|{\bf k}_\perp|^2$. { In Eq. \eqref{coupls3}, one has}     
\be
{\cal L}_{ij}={1\over 8\pi^2} ~{(\mu^2-\Lambda^2)^2  \over M^2 }
\int^1_0 dv~v^2~(1-v)^2 \int {dk^-\over 2 \pi}\nonu
\times ~
\Bigl\{ a^0_{ij}+ a^1_{ij}(v)~(p\cdot k)+a^2_{ij}(v)~(p\cdot k)^2+a^3_{ij}(v)~
k^2 \nonu
+ (1-v) \Bigl[(p\cdot k)^2-M^2 k^2 \Bigr]~\Bigl[d^0_{ij}+ d^1_{ij}~(p\cdot
k)\Bigr] \Bigr\}
\nonu \times ~{\cal S}(k^-,v,z,z',\gamma,\gamma')\label{l1}\ee
{In Eq. \eqref{l1}, the coefficients 
$a^\ell_{ik}(v)$ and $d^\ell_{ij}$ do not contain any dependence upon $k$ and
can be easily obtained from the  coefficients  $c_{ij}$ in Eq. \eqref{coupls1} }after singling out 
 the powers
of $k^\mu$ (recall  that a third power can be present,  at the most). This is
the  key ingredient  for correctly addressing the issue of the $k^-$ singularities.
 Moreover, the following definition has been adopted 
\be{\cal S}(k^-,v,z,z',\gamma,\gamma')=
{ 1\over
 \Bigl[(1-z)k^- + (1-z)k^-_d+i\epsilon\Bigr]}\nonu \times~
{1 \over\Bigl[(1+z)k^- -(1+z)k^-_u-i\epsilon\Bigr]
}
 \nonu
 \times
 ~{3
 k^-k^+_D+3\ell_D
 +F_v\over
 \Bigl[k^+_D k^-+\ell_D
+F_v+i\epsilon\Bigr]^3~
\Bigl[k^+_D k^-+\ell_D
+i\epsilon\Bigr]^2}
\ee    
with $F_v=(1-v)\Bigl ( \mu^2 -\Lambda^2\Bigr)$, $k^+_D=v(1-v)~(z'-z)M/2$ and 
\be
\ell_D=
-v(1-v) \Bigl(\gamma+zz'{M^2\over 4}-{z'}^2 {M^2\over 4}\Bigr)
\nonu-v\Bigl(\gamma'  +z'^2m^2+ (1-z'^2)\kappa^2\Bigr) -(1-v)\mu^2
~~, \nonu
k^-_{u(d)}= \pm{M\over 2}\mp{2 \over M(1\pm z)}(\gamma+m^2)~~.
\ee
It is easily seen   that the analytical integration on
$k^-$ of  \eqref{l1} involves integrals like
\be
{\cal C}_j
=
\int_{-\infty}^{\infty} {dk^-\over 2 \pi}
 (k^-)^j ~{\cal S}(k^-,v,z,z',\gamma,\gamma')
\label{cn}\ee
with $j=0,~1,~2,~3$, as dictated by the content in $k^\mu$ of $c_{ij}(k,k'',p)$.
For $k^+_D\ne 0$  and   $j\le 3$, one  can safely close 
the arc at infinity, in the
complex plane, and get the non singular contribution to ${\cal L}_{ij}$, namely
 the only part considered  in   Ref.
 \cite{CK2010} (i.e.  Eq. (18)).

  For describing a two-fermion system 
 or 
 for
 generalizing NIR to  massive vector
 constituents, one has to fully evaluate    ${\cal C}_j$, carefully analyzing 
 the case when 
  $k^+_D=0$. One can recognize through a simple counting rule that the tricky 
  powers are $j=2,3$, even if  $n>3$ is chosen in \eqref{nakaw}.
In  Ref. \cite{Yan},  singularities appearing  
in the infinite-momentum-frame quantum field theory
are investigated in details, singling out 
 the following singular 
integral, suitable for our purposes,
\be
{\cal I}(\beta,y)=\int_{-\infty}^\infty 
{ dx \over \Bigl[\beta x -y\mp i \epsilon\Bigr]^2}= \pm 
{2\pi i~\delta (\beta)\over \Bigl[-y\mp i \epsilon\Bigr]}
\label{yang1}\ee We also need  $(1/2)~\partial {\cal I}(\beta,y)/\partial y$,
  easily deduced from Eq. \eqref{yang1}.
Then, one gets our main result
(details in \cite{dFSV1}), namely
the singular contribution to ${\cal L}_{ij}$, given by
\be
{\cal L}^S_{ij} = -{i\over M}{1\over 8\pi^2} 
{(\mu^2-\Lambda^2)^2\over 2~(1-z^2)}~
\int^1_0 dv~v~(1-v)\nonu \times~
\Bigl\{{ \delta(z'-z)\over \Bigl(\tilde\ell_D
+F_v\Bigr)^2 ~\tilde\ell_D  } 
~\Bigl[a^2_{ij}(v)+(1-v)\Bigl(d^0_{ij}
+{M^2\over 4} z~d^1_{ij}  
\nonu+  \frac{2 z(\gamma +m^2)}{(1-z^2)} d^1_{ij}\Bigr)\Bigr] 
 + { d^1_{ij}\over v}~ \Bigl[{\partial \over \partial z'}\delta(z'-z)\Bigr]~
 {\cal D}^S_3\Bigr\}
\label{lsing}\ee
where  we used $\delta(x)/x=-d \delta(x)/d x$ and 
\be
\tilde \ell_D=-(1-v) ~(v\gamma+\mu^2)
-v\Bigl[\gamma'  +z^2m^2+ (1-z^2)\kappa^2\Bigr] 
\nonu
{\cal D}^S_3= 
{1\over F^2_v} ~
 \Bigl[{F_v
\over \ell_D
+F_v 
} + 
\ln\Bigl({ \ell_D \over  \ell_D
+F_v}\Bigr)  \Bigr]
\ee
The derivative of the Dirac delta-function is not an issue, since
 in our numerical method for solving the coupled integral equations   \eqref{coupls3}, 
after taking
into account Eqs. \eqref{coupls4}, \eqref{lsing}, and the non singular
contribution to ${\cal L}_{ij}$ 
we expand the Nakanishi weight functions $g_i(\gamma',z.;\kappa^2)$
 on a suitable basis. As in Ref. 
\cite{FSV2} for  two-scalar bound states, 
the basis is   composed by Laguerre and Gegenbauer polynomials
  (with the
 needed weights). It turns out that one can safely integrate 
$\partial\delta(z'-z)/\partial z'$ 
by part \cite{dFSV1}, given the smoothness of  our basis and the boundary 
property 
$g_i(\gamma',z'=\pm 1;\kappa^2)=0$. Then one can obtain
   an eigen-problem of the type $B~v=g^2 ~A~v$, (with $B$ and $A$  
 suitable  matrices). In our basis, we have  up to 
 $44$  Laguerre polynomials (with the same parameters as in Ref. \cite{FSV2}) and $44$ Gegenbauer ones, with indexes equal to $5/2,
7/2,7/2,7/2$ for 
 $g_i(\gamma',z.;\kappa^2)$ with  $i=1,2,3,4$, respectively. Moreover, 
 the small quantity to be added to $A_{ii}$ holds $\epsilon=10^{-7}$, and  
the number of Gaussian points is $120$, that becomes $180$  for analyzing the
case when the binding energy, 
in unit of $m$, $B/m=2-M/m$ is equal
to  $0.01$. 

In the studies of  BSE, it is customary to assign a
value to the binding energy  $B/m$,  and, in correspondence, look for  
an eigenvalue $g^2$.  If 
the eigenvalue
exists then the whole procedure is validated.
 Tables \ref{tab1} (scalar coupling) and   \ref{tab2} (pseudoscalar coupling)
 show the comparison between the values of $g^2$ 
obtained within our 
approach, where the singularities have been singled out and analytically 
evaluated,  
and both (i) the calculations  by Ref.
\cite{CK2010}, where  a non trivial numerical treatment of the 
singular behaviors was introduced (without  recognizing the possibility of a systematic
analysis of 
the singularities as in \cite{Yan}) and (ii)  the available numerical
results in Euclidean space \cite{Dork}, with a suitable number of digits. 
\begin{table}[t]
 \caption{The squared scalar coupling constant vs the binding energy 
 for  two masses of the
 exchanged particle $\mu/m=0.15$ and $\mu/m=0.50$. First column: binding energy.
 Second column: coupling constant $g^2$ for $\mu/m=0.15$,  obtained
 by taking analytically into account the fermionic singularities, (see text).
 Third column:  results for $\mu/m=0.15$, from Ref. \cite{CK2010} with
 a numerical treatment of  the singularities. Fourth column: the same as the second one,  but for
 $\mu/m=0.50$. Fifth column: the same  as the third one, but for
 $\mu/m=0.50$. Sixth column: 
 results in Euclidean space from Ref. \cite{Dork}. In the vertex
 form factor  it is taken $\Lambda=2$, as in \cite{CK2010} and
 \cite{Dork}.}  \label{tab1}
 \begin{center}
 \begin{tabular}{|c|c|c|}
 \hline
 ~~~~~~$~~~\mu/m=0.15~~~~$~~~~~~~&~&~~~~$~~~\mu/m=0.50~~$~~~~~~~~~~~~~~~\\
 \hline\end{tabular}
 \begin{tabular}{|c|c|c|c|c|c|c|}
 \hline
 B/m&$~g^2_{dFSV}$(full)~&~~~$g^2_{CK}~~$~&~&$~g^2_{dFSV}$(full)~&~~~$g^2_{CK}$~~~& ~~~~$g^2_{E}$~~~~\\
 \hline
 0.01&7.844 &7.813&~&25.327  &25.23&-\\
 0.02& 10.040&10.05&~&29.487& 29.49&-\\
 0.04&13.675 &13.69&~&36.183& 36.19&36.19\\
 0.05&15.336 &15.35&~&39.178 & 39.19&39.18\\
 0.10&23.122&23.12&~&52.817 & 52.82&-\\
 0.20&38.324 &38.32&~&78.259  &78.25&-\\
 0.40&71.060 &71.07&~&130.177  &130.7&130.3\\
 0.50& 88.964 &86.95&~&157.419   &157.4&157.5\\
 1.00& 187.855  & - &~&295.61  &- & - \\
 1.40& 254.483 & -&~&379.48 & -&-\\
 1.80& 288.31  & -&~&421.05 &- &-\\
 \hline
 \end{tabular}
 \end{center}
 \end{table}
  \begin{table} 
 \caption{The same as in Table \ref{tab1}, but  for a
 pseudoscalar coupling. }\label{tab2}
 \begin{center}
 \begin{tabular}{|c|c|c|}
 \hline
 ~~~~~$~~\mu/m=0.15$~~~~~~~~~~~~~~&~&~~~$~~\mu/m=0.50$~~~~~~~~~\\
 \hline\end{tabular}
 \begin{tabular}{|c|c|c|c|c|c|c|}
 \hline
 B/m&$~g^2_{dFSV}$(full)~ & ~~~$g^2_{CK}$~~~&~&$~g^2_{dFSV}~$(full) 
 & ~~~$g^2_{CK}$~~~\\
 \hline
 0.01&  225.7 &224.8&~& 422.6 & 422.3\\
 0.02&  233.2& 232.9&~&430.5 & 430.1\\ 
 0.04&  243.1& 243.1&~&440.9 & 440.4\\
 0.05&  247.1 & 247.0&~&444.9 & 444.3\\
 0.10&  262.1 &262.1&~&460.4 & 459.9\\
 0.20&  282.9 &282.9&~&482.1 & 480.7\\
 0.40&  311.7 &311.8&~&513.3 & 515.2\\
 0.50&  322.9 &323.1&~&525.8 & 525.9\\
 1.00&   362.3 & - &~&570.9 &  -\\  
 1.40&  380.1 & -&~&591.8 & -\\
 1.80&  388.7  & -&~& 602.1 &-\\
 \hline
 \end{tabular}
 \end{center}
 \end{table}

\begin{figure}
 \hspace{-0.5cm}\includegraphics[width=10.cm] {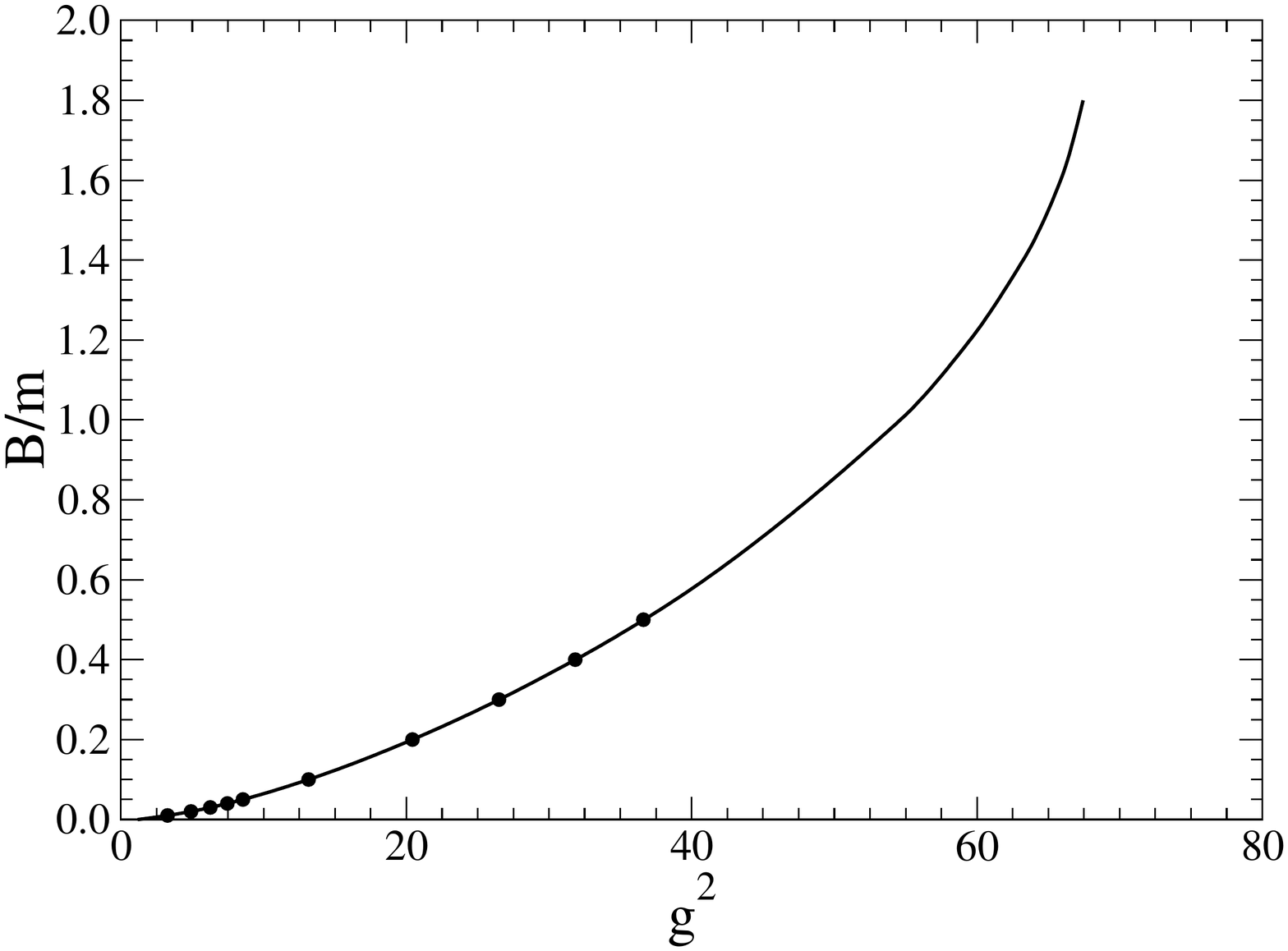}

 \caption{The binding energy $B/m$ vs $g^2$  for a
  massless vector exchange. Solid  line: our
 calculations, with the exact analytical treatment of the end-point singularities.
 Full dots: $g^2$ from Ref. \cite{CK2010}, with a numerical
  treatment of the singularities.
 A critical value $g_{crit} $ is clearly approached for $B/m \to 2$.
  } \label{fig1}
  \end{figure}

 Notably,  
we were also able to extend our calculation up to $B/m\sim 2$, 
namely when the expected critical behavior of a
$\phi^3$ theory manifests itself \cite{Baym}, i.e. 
where $\partial B/\partial g^2 \to \infty$.  This is well illustrated in Fig.
\ref{fig1}, where    the comparison between our
calculations for the vector coupling and the ones by \cite{CK2010} is also shown.

 { The achieved full agreement, within the adopted numerical accuracy,
strongly supports the validity of our analytical method for treating
 the
singularities that plague  ladder BSE, when an interacting
two-fermion system is considered. The most severe
singularity is met when the third power of $k^-$ appears in the numerator of 
the
 kernel in Eq. \eqref{l1}. The powers of $k^-$ are  generated  only by
 the external propagators
 and the
structure of the BS amplitude, present in the lhs of \eqref{bse}.
Therefore  the 
 highest
 power of $k^-$ { is fully independent of the kernel complexity.  
 For instance},
 in the case of a two-vector system, this simple counting rule leads to expect 
 derivatives of the Dirac delta-function not too
high ($\ge 2$, depending only upon the complexity of the BS amplitude, like in Eq.
\eqref{bsa}), and therefore 
still manageable within our
approach.}

\begin{figure}
\includegraphics[width=10.cm] {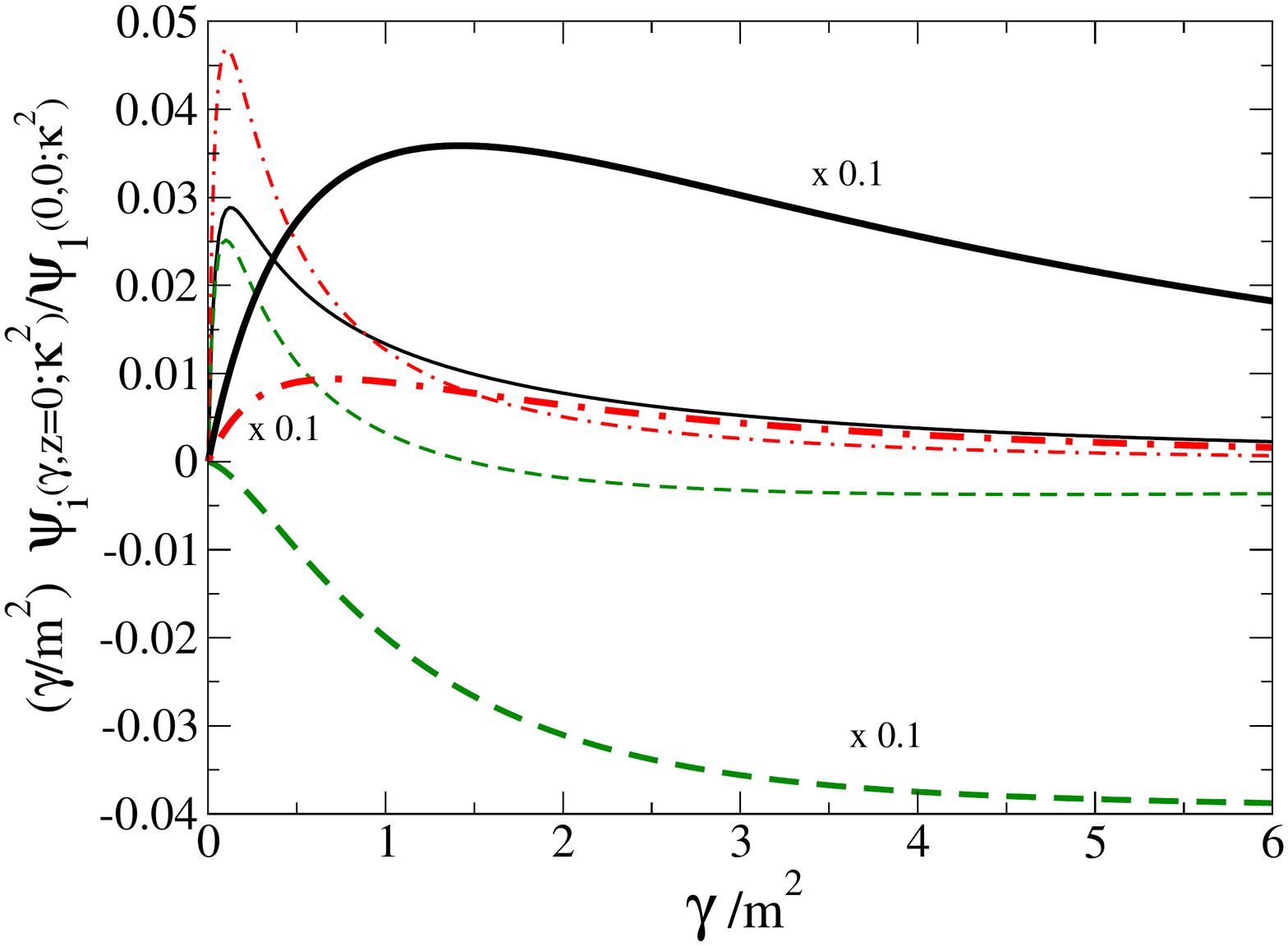}

 \caption{The light-front amplitude $\psi_i$ times $\gamma/m^2$ (cf Eq. \eqref {coupls4})  vs 
 $\gamma/m^2$ at fixed $z=0$, 
  for the vector coupling and $B/m=0.1$ (thin lines) and $1.0$ (thick lines).
Solid line: $(\gamma/m^2)~\psi_1$. Dashed line:  $(\gamma/m^2)~\psi_2$. 
Dot-dashed line:  $(\gamma/m^2)~\psi_4$. Notice that $\psi_3=0$ for 
$z=0$ (see text).
The lines for  $B/m=1.0$ have been divide by 10.}\label{fig2}
  \end{figure}

{ In Fig. \ref{fig2},  the LF amplitude $\psi_i(\gamma,z=0;\kappa^2)$ 
(cf Eq. \eqref {coupls4}) times a factor $\gamma/m^2$ are shown for the 
{\em vector coupling}, with   
$B/m=0.1$ and $ 1$ (i.e.  weak and  strong regimes, respectively). 
For $z=0$, $\psi_3$ vanishes, since it is odd  in
$z$. Figure \ref{fig2} puts in evidence the power-like tails
of $\psi_i$, as expected for a hadronic system by a simple counting 
rule \cite{Ji2003} that predicts $1$ for 
 the fall-off power of the pion valence wave function. { Such a power $1$
 is a distinctive feature of the ladder kernel triggering the high-momentum
 tail and the spin 1/2 (for  scalars,  one has power $2$ \cite{FSV2}).}
Notice that the LF amplitudes  (see Refs. \cite{FSV2,dFSV1})
 are  basic ingredients for non perturbative evaluations of 
 valence wave functions and  momentum distributions, in the physical space.}

  The robustness of the technique  based on 
  NIR for solving 
the BSE with spin degrees of freedom encourages to extend 
this novel tool to many areas, since  old limitations 
 constraining the calculations to
an unphysical
space can be removed. The  approach  can deal with further
dynamical effects, since the analytical structure 
of BS kernels,  truncated at any power of the coupling constant,  
 is made explicit as in the ladder case (see, e.g.,
\cite{FSV1} for the half-off-shell T-matrix),  allowing the LF projection. 
  
\begin{acknowledgments}
We gratefully thank  J.
Carbonell and V. Karmanov for very stimulating  discussions.
TF and WdP acknowledge the warm hospitality of INFN Sezione di Roma and  thank 
 the partial financial 
support from the Brazilian Institutions:
 CNPq, CAPES and 
FAPESP. 
GS  thanks the partial support of CAPES
 and  acknowledges the warm hospitality of  the  Instituto Tecnol\'ogico de 
 Aeron\'autica.
\end{acknowledgments}


\begin{thebibliography}{99}
\bibitem{BS51}E. E. Salpeter and  H. A. Bethe,  Phys. Rev. {\bf 84}, 1232 (1951).
\bibitem{FSV1}T. Frederico, G. Salm\`e and M.  Viviani, 
  Phys. Rev. {\bf D 85}, 036009 (2012). 

\bibitem{Kusaka} K. Kusaka, K. Simpson, and A. G. Williams,  Phys. Rev. {\bf D 56}, 5071
(1997).

\bibitem{CK2006} V. A. Karmanov, J. Carbonell, 
Eur. Phys. Jou. {\bf A 27}, 1 (2006).

\bibitem{FSV2}T. Frederico, G. Salm\`e and M. Viviani, 
  Phys. Rev. {\bf D 89}, 016010 (2014).
\bibitem{Tomio2016} C. Gutierrez, V. Gigante, T. Frederico, G. Salm\`e, M. Viviani and L.
Tomio,  Phys. Lett. {\bf B 759},
131(2016).
\bibitem{FSV3}
T. Frederico, G. Salm\`e and M. Viviani,   	Eur. Phys. J. {\bf C 75}, 398 (2015). 


\bibitem{CK2010}
  J.~Carbonell and V.~A.~Karmanov,
  Eur.\ Phys.\ J.\ A {\bf 46}, 387 (2010).
  
 
 
 \bibitem{CK2006b} J. Carbonell, V. A. Karmanov, 
Eur. Phys. Jou. {\bf A 27}, 11 (2006). 
 
\bibitem{Dork} S. M. Dorkin, M. Beyer,  S. S. Semikh,  and L. P. Kaptari, 
 Few-Body Sys. {\bf  42} 1 (2008), and private communication.
\bibitem{Ji2003} X. Ji, J.P. Ma and F. Yuan, Phys. Rev,. Lett. {\bf 90}, 241601
(2003).
\bibitem{dFSV1} W. de Paula, T. Frederico, R. Pimentel, G. Salm\`e and M. Viviani, in
preparation.
\bibitem{Yan} T.M. Yan ,
 Phys. Rev. {\bf D 7}, 1780 (1973).


\bibitem{Baym} G. Baym,  Phys. Rev. {\bf 117}, 886 (1960).

\end{thebibliography}
\end{document}